\documentstyle[graphicx,12pt]{article}

\begin{document}
\begin{titlepage}

\vspace{1in}

\begin{center}
\Large{\bf Inflation and Braneworlds}

\vspace{1in}

\normalsize

\large{James E. Lidsey$^1$}

\normalsize
\vspace{.7in}

{\em Astronomy Unit, School of Mathematical 
Sciences,  \\ 
Queen Mary, University of London, Mile End Road, LONDON, E1 4NS, U.K. \\}

\end{center}

\vspace{1in}

\baselineskip=24pt
\begin{abstract}
\noindent An introductory review of the Randall-Sundrum type II braneworld 
scenario is presented, with emphasis on 
the relationship between the density and 
gravitational wave perturbations
that are generated during inflation. The implications 
of relaxing the reflection symmetry in the fifth 
dimension are considered. The effects of 
including a Gauss-Bonnet combination of higher-order curvature 
invariants in the bulk action are briefly discussed. 
\end{abstract}

\vspace{.7in}
\noindent {\em Invited plenary talk presented at the 
Fifth Mexican School of Gravitation and Mathematical 
Physics: 
`The Early Universe and Observational Cosmology', Playa del Carmen, 
November, 2002.}

\vspace{.7in}

\noindent $^1$Electronic mail: J.E.Lidsey@qmul.ac.uk

\end{titlepage}

\section{Introduction}

A unified description of the origin and very early 
evolution of the universe that is consistent both
with our understanding of unified field theory and 
astrophysical observations is one of the primary goals of 
particle cosmology. A synthesis of these two disciplines 
provides a unique window to high energy physics that would 
otherwise be inaccessible to any form of terrestrial experiment. 

From the observational side, 
recent years have witnessed rapid advances in the quality and availability
of high precision data from numerous cosmic microwave background (CMB) 
and high redshift surveys. This is resulting in ever more stringent 
constraints on models of the early universe and the trend
is certain to continue 
in light of the anticipated data that will become available in 
the near future. Specifically, recent measurements 
from the Wilkinson Microwave Anisotropy Probe (WMAP) \cite{temp}
are entirely consistent with a universe that has a total density 
that is very close to the critical density, implying that the 
curvature of the universe is very close to spatial flatness 
\cite{wmap}. 
On the other hand, there is by now considerable evidence
from a variety of sources --  including the CMB power spectrum, galaxy 
clustering statistics, peculiar velocities, the baryon mass fraction in galaxy 
clusters and Lyman--$\alpha$ forest data -- that the density of 
clumped baryon and non--baryonic matter can be no more than 
30$\%$ of the critical density. Moreover, spectral and photometric 
data from high redshift surveys of type Ia supernovae 
\cite{Perlmutter}
indicate that the expansion of the universe may be accelerating at the 
present epoch, thereby requiring the existence of some form of exotic 
`dark energy' or `quintessence' field that 
contributes the remaining 70$\%$ of the total energy density.

The popular explanation of this diverse set of observations 
is the {\em inflationary scenario} \cite{simplest}, 
whereby the universe underwent an 
epoch of very rapid, accelerated expansion sometime before  
the electroweak phase transition. Inflation is presently 
the cornerstone of modern, early universe cosmology. 
(For a review, see, e.g., Ref. \cite{lidlyth}). Not only is inflation able to 
resolve the horizon and flatness problems of the hot, big 
bang model, it also provides the mechanism for generating 
the primordial density perturbations necessary for 
galaxy formation \cite{perturbations}. 
In the simplest class of inflationary models, the 
accelerated expansion is driven by the potential energy arising 
through the self--interactions of a single quantum 
scalar field, referred to as the {\em inflaton} field and denoted $\phi$. 
If the potential is sufficiently flat and smooth, the field is able to 
slowly roll towards the minimum of its potential. In this case, the 
kinetic energy of the field is subdominant and its pressure 
becomes sufficiently negative for
the strong energy condition of General Relativity to become violated. 
Inflation ends as the 
field reaches its ground state and the hot big bang model 
is recovered through a reheating process.

It is now widely believed that the observed large--scale 
structure in the universe evolved through the process of 
gravitational instability from density perturbations 
that were generated {\em quantum mechanically}  during 
the inflationary expansion. In single field inflationary models, 
the perturbations are predicted to be adiabatic, nearly scale--invariant and 
Gaussian distributed. Moreover, inflation results in an effectively 
flat universe. The current CMB data, most notably 
from WMAP \cite{wmap,wmap1,kogut,komatsu}, supports these predictions 
whilst simultaneously providing strong constraints on such models 
\cite{bridle,kkmr}. In particular, 
an anti--correlation between the temperature and polarization E--mode maps
of the CMB on degree scales 
has been detected by WMAP \cite{kogut}, 
thereby providing strong evidence 
for correlations on length scales beyond the Hubble radius \cite{white}.

Despite the success of inflationary cosmology in passing these key 
observational tests, there is presently no canonical theory for 
explaining the origin of the inflaton field. Consequently, it 
is imperative to establish that inflation can arise generically 
within the context of unified field theory. Superstring theory has 
emerged as the leading candidate for such a theory of the 
fundamental interactions, including gravity. 
Developments over recent years towards a non--perturbative formulation of 
the theory have indicated that the five, anomaly--free, 
supersymmetric perturbative string theories -- known respectively as 
types I, IIA, IIB, SO(32) heterotic and ${\rm E}_8 \times {\rm E}_8$ 
heterotic -- represent different 
limits of a more fundamental theory referred to 
as M--theory. (For reviews, see, e.g., Refs. 
\cite{string,Dbrane}). M--theory was originally defined as the strong 
coupling limit of the type IIA superstring \cite{witten}. However, 
since its infra--red (low--energy) 
limit is eleven--dimensional supergravity, 
it must be more than another theory of superstrings \cite{witten,townsend}. 

Given this change of perspective, it is crucial to study the cosmological 
consequences of string/M--theory \cite{lwc}. Supersymmetry implies that 
a consistent quantum string theory can only be formulated 
if spacetime is higher--dimensional. 
Given that these extra dimensions are not observed, 
some mechanism is required to ensure that they remain 
undetected. One possibility 
is that the dimensions are compactified through the Kaluza--Klein mechanism. 
In this case, tests of quantum electrodynamics 
limit the size of the extra dimensions to be less than $10^{-17}
\, {\rm cm}$. 

On the other hand, a key theoretical development 
has been the realization that the standard model fields 
(quarks, electrons, photons, etc.)
may be confined to a four--dimensional domain wall or `membrane' 
that is embedded in a higher--dimensional space (referred to as the bulk).
This picture has developed following the  
discovery that the quantum dynamics of D--branes can be 
described by open strings whose ends are fixed on the brane \cite{polchinski}. 
In string theory, branes are static, solitonic configurations 
extending over a number of spatial, tangential dimensions. 
Thus, a $0$--brane may be viewed as a pointlike particle or a black hole, a 
$1$--brane represents a string, a $2$--brane a membrane, and so forth. 
In this picture, our observable, four--dimensional universe 
is interpreted as a $3$--brane. The spatial dimensions tangential to 
our 3--brane describe our familiar three--dimensional space of length, 
width and height. The only long--range interaction that propagates 
in the bulk dimensions 
is gravitational. In this case, corrections 
to Newton gravity necessarily arise,
but the weak nature of gravity implies that any modifications 
can not presently be observed below scales of $1 \, {\rm mm}$. 
In principle, therefore, the 
extra dimensions may extend over scales that are many orders of magnitude  
larger than previously thought possible 
and, depending on the model, may even be infinite 
in extent. This paradigm shift in our understanding of 
the observable universe is referred to 
as the {\em braneworld scenario}.

The radical proposal, therefore, is that {\em our universe is a brane 
embedded in a higher--dimensional space}. 
The implications for cosmology, and for our 
understanding of the inflationary scenario in particular, 
are significant and there is presently a high level of active 
research in this field. Broadly speaking, the key objectives from the 
astrophysical and cosmological perspectives are: 

$\bullet$ To determine the nature of cosmological solutions that 
are possible in braneworld scenarios, to investigate their asymptotic 
early-- and late--time behaviours, and to uncover important differences 
and similarities between braneworld scenarios and conventional 
cosmologies based on Einstein gravity. 

$\bullet$ To establish the conditions whereby inflation may occur, both
in the arena of the early universe and at the present epoch (quintessence
scenarios) and to determine whether inflation is more or 
less generic in this new paradigm. 

$\bullet$ To investigate the production of scalar (density), 
vector (electromagnetic) and tensor (gravitational wave) perturbations 
during braneworld inflation.

$\bullet$ To develop cosmological tests of inflationary braneworld 
scenarios and determine whether the perturbations generated are compatible 
with limits imposed by the CMB power spectrum and large--scale 
structure observations. 

Ultimately, such a programme will yield unique information 
on the dimensionality of the universe.

\section{Types of Braneworlds}

It is impossible in a talk of this nature to
fully review the vast body of work in this field. 
The emphasis from a cosmological point of view 
has focused on models consisting of a single brane or 
of two or more parallel branes. (For early papers 
see, e.g., Refs. \cite{early}). From an historical 
point of view, a significant development was the 
interpretation by Ho\v{r}ava and Witten
of the strongly coupled limit of the 
${\rm E}_8 \times {\rm E}_8$ heterotic string 
as M--theory compactified 
on the eleven--dimensional orbifold $R^{10} \times S^1/Z^2$ \cite{hw}. The 
weakly coupled limit of this string theory then corresponds to the 
limit where the radius of the circle (as parametrized by the 
value of the dilaton field) tends to zero. 
The orbifold $S^1/Z_2$ may be viewed 
as the segment of the real line bounded by 
two fixed points on the circle, such that the orientation 
of the circle is reversed by the $Z_2$ 
transformation, $y \rightarrow -y$. Gravitational anomalies are 
cancelled by placing the two sets of ${\rm E}_8$ gauge supermultiplets 
on each of the ten--dimensional orbifold fixed planes. 
An effective five--dimensional theory may then be 
derived by compactifying on an appropriate Calabi--Yau surface \cite{lukas}. 

\begin{figure}
\centering
\includegraphics[height=8cm]{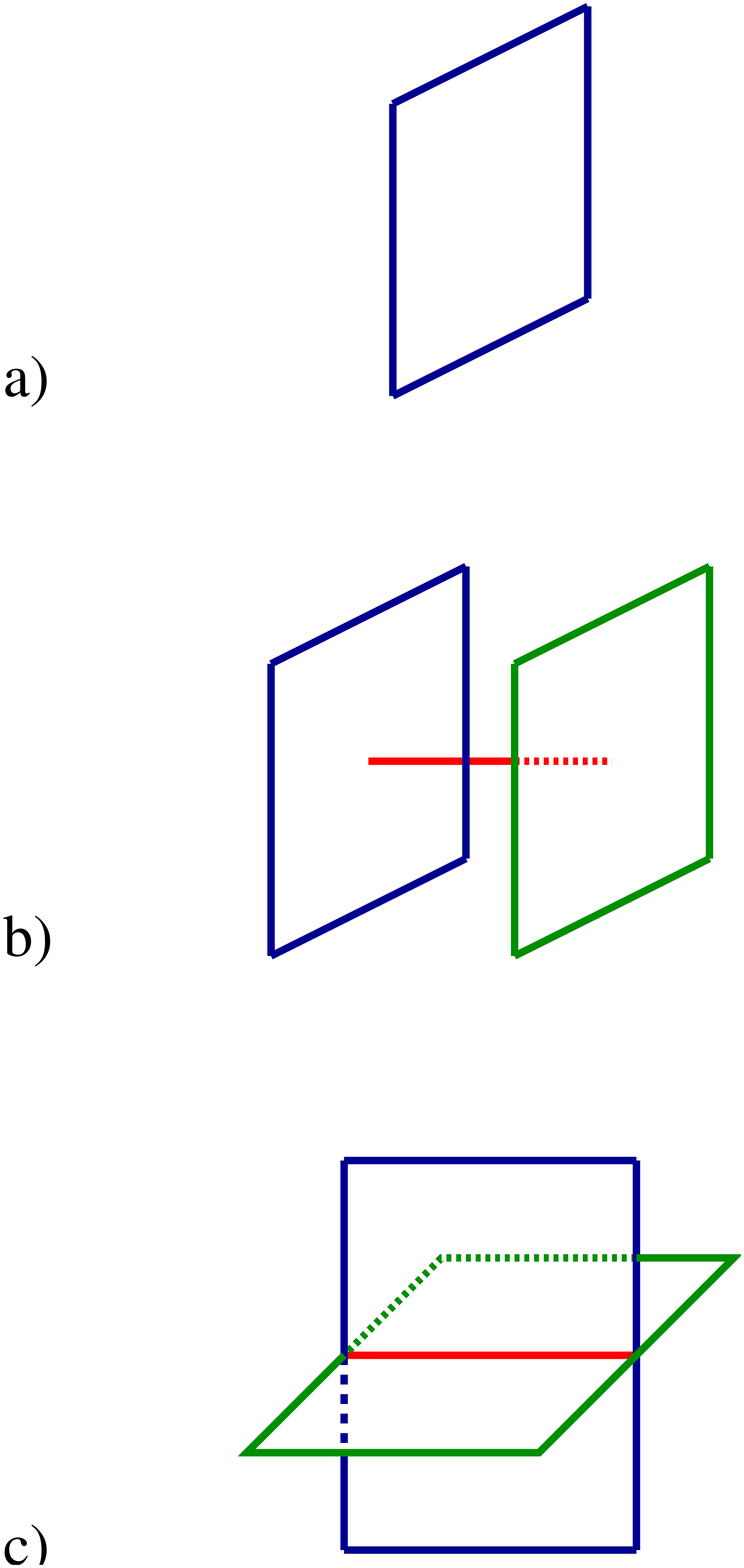}
\caption{Illustrating possible braneworld configurations: 
(a) a single brane embedded in a higher--dimensional space; (b) two 
parallel branes with equal and opposite tension; (c) intersecting branes.}
\label{mexico1}       
\end{figure}

Cosmological solutions admitted in this theory 
were found and analyzed \cite{hwcos}. In particular, 
models where the branes approach and move away from each other 
were found and interpreted in terms of the 
pre--big bang scenario, an earlier string--inspired inflationary 
scenario driven by the kinetic energy of a scalar field \cite{ven}. 
Recently, the idea of interpreting the 
big bang in terms of brane collisions has been advocated 
through the `ekpyrotic' scenario \cite{turok}. In these models, 
the brane dynamics is reduced to an effective four--dimensional 
theory, where a scalar field parametrizes the brane separation. 

A further key development was 
the proposal that the hierarchy problem of particle physics (namely the 
problem of understanding why the weak scale is so much smaller than the 
Planck scale) could  be alleviated if the volume of the extra dimensions 
were to be made sufficiently large \cite{hier}. 
In general, the four--dimensional Planck scale, $m_4$, 
is related to the $(4+n)$--dimensional Planck scale, $M$, through the 
relationship $m_4^2 = M^{n+2}V_n$, where $V_n$ 
is the volume of the compact space. 
In the model of Ref. \cite{hier}, the extra dimensions were assumed to be 
topologically equivalent to a $n$--torus and a single brane configuration 
was considered.  
Pursuing such ideas further, Randall and Sundrum considered 
two parallel branes, with equal and 
opposite tension, embedded in five--dimensional Anti--de Sitter (AdS)
space, with a $Z_2$ reflection symmetry imposed in 
the fifth dimension \cite{RSI}. In this model, the weak scale is generated 
from a larger (Planck) scale through an exponential hierarchy 
arising directly from the five--dimensional AdS geometry. 

Perhaps of more interest from a cosmological viewpoint is the second 
Randall--Sundrum `type II' (RSII)
scenario, 
consisting of a single brane embedded in five--dimensional AdS 
space \cite{RSII}. 
Formally, this model may be interpreted in terms of the 
first Randall--Sundrum model, where the negative tension brane 
is taken to infinity. The bulk space may also contain a black hole. 
In this case it corresponds to the five--dimensional Schwarzschild--AdS 
solution. The RSII model is interesting because it 
is  simple enough for analytical results to be derived, 
yet is sufficiently rich for new physics to be uncovered. 
Indeed, one of the key areas of interest in theoretical physics at present is 
focused towards the `holographic principle' \cite{holographic}. 
In short, the holographic 
principle implies that the number of degrees of 
freedom associated with gravitational dynamics is determined 
by the boundary of the spacetime rather than by its volume. This 
follows naturally from the idea that the state of maximal entropy for 
a given volume, $V$, is determined by the size of the largest black hole 
that can be contained within $V$ \cite{holographic}. 
The AdS/Conformal Field Theory (CFT) correspondence provides 
a realization of this principle within the context of 
string theory by establishing a duality 
between semi--classical $d$--dimensional gravity in AdS space and a quantum 
CFT located on its boundary \cite{adscft}. It has 
recently become apparent that the AdS/CFT correspondence 
is closely related to braneworld cosmology and the RSII scenario  
in particular \cite{hrr}. More specifically, one may view the RSII braneworld 
as being dual to a CFT (with an ultra--violet cut off) that is coupled to 
gravity on the brane.  Remarkably, when one identifies the 
entropy, mass and Hawking temperature of the AdS black hole with 
the entropy, energy and temperature of the CFT, it is found that the Cardy
entropy formula of the CFT coincides precisely with the 
Friedmann equation when the brane passes the 
black hole event horizon \cite{ver}. 

We focus on the RSII scenario in the remainder of this talk. 
Before proceeding, however, it is worth highlighting 
an alternative class of `intersecting' braneworlds. 
Configurations in supergravity theories that describe the intersection of 
two or more $p$--branes have played a prominent role 
in advances in string theory \cite{argurio}. 
A $p$--brane can be supported, for example, when the components 
of the antisymmetric form fields that arise in the string and M--theory 
effective actions have non--trivial flux over the 
compactifying manifold.  
Depending on the degree of the form field and the 
nature of the internal space, such a model may represent the 
intersection of two higher--dimensional branes (See Fig. \ref{mexico1}c).
One specific example is the intersection of two $5$--branes over a 
$3$--brane \cite{argurio}. In this picture, our 
observable four--dimensional universe corresponds to the 
intersection of the branes. Solutions representing curved (time--dependent),  
intersecting domain walls were recently found \cite{lidsey1}. 

\section{The Randall--Sundrum Type II Braneworld}

In the Randall--Sundrum type II 
scenario, a co--dimension one brane is embedded 
in five--dimensional AdS space with a $Z_2$ reflection symmetry
imposed on the bulk. The action is given by 
\begin{eqnarray}
\label{RSaction}
S=\int_{\cal{M}}   d^5x \sqrt{G} \left[ 2M^3 \hat{R} -\Lambda \right] 
+ \int_{\partial {\cal{M}}} d^4x \sqrt{h} \left( \lambda +L_{\rm 
matter} \right) ,
\end{eqnarray}
where $\hat{R}$ is the Ricci curvature scalar of the 
bulk spacetime, ${\cal{M}}$, 
with metric $G_{AB}$, $\Lambda$ is the five--dimensional
(negative) cosmological constant, $G \equiv {\rm det} {G}_{AB}$, 
$h$ is the determinant of the metric induced on the boundary of 
${\cal{M}}$, 
$\lambda$ is the tension of the brane, $L_{\rm matter}$ is the 
Lagrangian density of the matter on the brane and $M$ is the 
five--dimensional Planck mass. 

The bulk solution has a metric of the form \cite{RSII}
\begin{equation}
\label{bulkmetric}
ds^2 = e^{ -2k |y|} \left( -dt^2 + dx^2_3 \right) 
+dy^2  
\end{equation}
for a given constant, $k$. 
This geometry is non--factorizable due to the presence of the 
exponential warp factor, in contrast to the standard Kaluza--Klein 
compactification 
schemes based on the periodic boundary conditions. Consequently, 
the fifth dimension $(y)$ may extend to infinity. Imposing 
four--dimensional Poincar\'e invariance on the brane world--volume
such that the metric corresponds to flat Minkowski spacetime 
requires the bulk cosmological constant to be fine--tuned with the 
brane tension \cite{RSII}:
\begin{equation}
\label{RStune}
\Lambda =-24 M^3 k^2 , \qquad \lambda = 24M^3 k  .
\end{equation}

The remarkable feature of this model is that 
the graviton equation of motion admits 
a zero energy ground state solution
that is localized around the domain wall.
This ground state is naturally interpreted (by a four--dimensional observer) 
as the four--dimensional, massless spin--2 graviton. 
A continuum of massive states also arises in the spectrum and these 
lead to corrections to the form of the Newton
potential. However, these corrections fall as the cube of the 
distance, $r$ \cite{RSII}: 
\begin{eqnarray} 
\label{Newtonpotential}
V(r) \approx \frac{G_N m_1 m_2}{r} \left( 1 + \frac{1}{k^2r^2} 
\right)  
\end{eqnarray}
and, if the warping of the bulk geometry is 
sufficiently  strong (i.e. the constant $k$ is sufficiently 
large), these massive states are suppressed near the 
brane and are therefore harmless. 
This indicates that  
{\em the curvature of the five--dimensional world 
effectively determines the four--dimensional physics}.

The cosmology of the RSII scenario arises 
due to the motion of the brane through the bulk space. 
An observer confined to the surface of the brane 
interprets such motion in terms of cosmic expansion or contraction
\cite{othermoves,reall,bv}. 
The `Friedmann' equation describing the cosmic dynamics may be derived 
within the context of the  
thin wall formalism of (five--dimensional) General Relativity. Since we are 
interested primarily in late--time inflationary 
dynamics, we focus on the simplest case where the 
world--volume of the brane corresponds to the spatially 
flat, Friedmann--Robertson--Walker (FRW) metric and consider 
a pure AdS bulk. The effect of a bulk black hole on the four--dimensional 
brane dynamics is formally equivalent to that of a relativistic 
perfect fluid contribution 
to the energy--momentum tensor and so is rapidly redshifted away by the 
accelerated motion of the brane. 

It is convenient to work with the five--dimensional metric expressed in 
static coordinates: 
\begin{eqnarray}
\label{staticbulk}
ds^2 =G_{AB} dx^Adx^B =-\frac{r^2}{L^2}dt^2 +\frac{L^2}{r^2}
dr^2 +r^2 dE^2_3  ,
\end{eqnarray}
where the constant $L$ is related to the bulk cosmological constant. 
The induced metric on the wall then has the desired form: 
\begin{eqnarray}
\label{inducedmetric}
h_{AB} =G_{AB} +n_An_B \nonumber \\
ds_4^2 =-d\tau^2 +a^2(\tau ) dE^2_3 ,
\end{eqnarray}
where $n^A$ is the unit normal vector to the brane. Cosmic
time  as measured on the brane is parametrized by $\tau$,
defined such that  
\begin{equation}
\label{cosmictime}
d\tau^2 =\frac{r^2}{L^2} dt^2 -\frac{L^2}{r^2} dr^2 
\end{equation}
and the radial coordinate of the brane in the bulk space 
determines the scale factor, $a(\tau )$:
\begin{equation}
\label{radial}
r=r \left[ a(\tau ) \right]  .
\end{equation}

The effective Friedmann equation is then 
derived directly from the Israel junction conditions \cite{israel}: 
\begin{equation}
\label{israeltwo}
K_{AB} =-4\pi G_5 \left( T_{AB} -\frac{1}{3} T h_{AB} \right) ,
\end{equation}
where $G_5$ is the five--dimensional Newton constant. 
These conditions relate the energy--momentum  tensor, $T_{AB},$ 
of the matter confined on the brane directly to  
the brane's extrinsic curvature, 
$K_{AB} \equiv {h^C}_{(A}{h_{B)}}^D \nabla_C n_D$.
Note that we have taken into account the 
${\rm Z}_2$ symmetry in this expression and that $T\equiv T^A_A$ 
is the trace of the energy--momentum.  

Conservation of energy--momentum on the brane
then follows as a direct consequence of the 
Codazzi equation:
\begin{equation}
\label{codazzi} 
\nabla_B K^B_A -\nabla_A K = 
\hat{R}_{BC} G^B_A n^C   .
\end{equation} 
It is straightforward to verify that for the 
case of a pure AdS bulk geometry, the
right--hand side of Eq. (\ref{codazzi}) is identically zero. 
Thus, substitution of the 
Israel junction conditions (\ref{israeltwo}) 
into Eq. (\ref{codazzi}) implies 
conservation of energy--momentum on the brane: 
\begin{equation}
\label{emconserve}
{^{(4)}} \nabla_{\mu} T^{\mu\nu} =0  .
\end{equation}
We will further assume that the 
energy--momentum tensor of the matter on the brane 
is given by the perfect fluid
form 
\begin{equation}
T^A_B |_{\rm brane} = \delta (y) {\rm diag} 
(-\rho , p, p, p, 0) ,
\end{equation}   
where $\rho$ and 
$p$ represent the energy density and pressure, 
respectively. Hence, we recover the standard expression of 
conventional cosmology: 
\begin{equation}
\label{fluid}
\dot{\rho} +3\frac{\dot{a}}{a} (\rho + p ) =0  ,
\end{equation}
where a dot denotes $d/d\tau$. 

The spatial components of the Israel junction conditions 
(\ref{israeltwo}), as given by 
\begin{eqnarray}
\label{spatialcomp}
K_{ij} =- \sqrt{\frac{1}{L^2} + \frac{\dot{a}^2}{a^2}} \delta_{ij}  ,
\end{eqnarray}
are now sufficient to derive the
effective Friedmann equation. (The time--time 
components of Eq. (\ref{israeltwo}) provide no new information 
for the model we are considering). We therefore deduce that 
\cite{bine,flanagan,4a,othermoves,bv,shiromizu}
\begin{eqnarray}
\label{dota}
\frac{\dot{a}^2}{a^2} = \left( \frac{4\pi G_5\rho}{3} \right)^2 
-\frac{1}{L^2}  .
\end{eqnarray}

Although the quadratic dependence of the Friedmann equation (\ref{dota}) 
may appear to be inconsistent with the Hubble expansion, 
and particularly with constraints form primordial nucleosynthesis, 
we must recall that the vacuum brane has a tension, $\lambda$. 
This implies that the total matter content on the brane can 
effectively be separated into two components, the dynamical 
matter, $\rho_B$, and the tension. Substituting into 
Eq. (\ref{dota}) then implies that 
\begin{eqnarray}
\label{quad}
H^2 =\frac{8\pi G_4}{3} \rho_B \left( 1+
\frac{\rho_B}{2\lambda} \right) + \left( 
\frac{4\pi G_5 \lambda}{3} \right)^2 -\frac{1}{L^2} ,
\end{eqnarray}
where $G_4 \equiv 4\pi \lambda G_5^2/3$. The constant terms 
are then cancelled by imposing the fine--tuning condition 
(\ref{RStune}), resulting in a Friedmann equation of the form
\cite{bine,flanagan,4a,othermoves,bv,shiromizu}
\begin{equation}
\label{branefriedmann}
H^2 = \frac{8\pi}{3m_4^2} \rho \left[ 1+\frac{\rho}{2\lambda} 
\right]  ,
\end{equation}
where the subscript `$B$' is dropped for notational simplicity and we
define the four--dimensional Planck mass, $m_4 \equiv G^{-1/2}_4$. 
The standard form of the Friedmann equation is recovered at 
low energy scales, $\rho \ll \lambda$, whereas the 
dependence on the energy density is modified 
to a quadratic form at high energies, $\rho \gg \lambda$. 

We now consider the implications of this term for inflationary
cosmology. 

\section{Braneworld Inflation}

\subsection{Scalar Field Dynamics}

Eqs. (\ref{fluid}) and (\ref{branefriedmann}) 
are sufficient to fully determine the cosmic dynamics 
on the brane once an equation of state has been specified 
for the matter sources. In what follows, we assume that the brane matter 
consists of a single scalar field that is confined 
to the brane and is self--interacting through 
a potential, $V(\phi )$. The conservation equation (\ref{fluid})
then implies that
\begin{equation}
\label{branescalar}
\ddot{\phi} +3H \dot{\phi} +V'=0 ,
\end{equation}
where a prime denotes differentiation with respect to the scalar 
field. We further 
assume the slow--roll approximation, $\dot{\phi}^2 \ll V$ 
and $|\ddot{\phi}| \ll H |\dot{\phi}|$. Eq. (\ref{branescalar})
then simplifies to $3H\dot{\phi}  \approx -V'$.

The slow--roll parameters, $\epsilon \equiv -\dot{H}/H^2$ and 
$\eta \equiv V''/(3H^2)$, may then be written in the form \cite{maartens}
\begin{eqnarray}
\label{epsilon}
\epsilon  \simeq \frac{m_4^2}{4 \pi} \, 
\left( \frac{V'}{V} \right)^2 \,
\left[ \frac{1+V/\lambda}{\left(2 + V/\lambda \right)^2} \right] \\
\label{eta}
\eta \simeq \frac{m_4^2}{8\pi} \left( \frac{V''}{V} \right) 
\left[ \frac{2\lambda}{2\lambda +V} \right]
\end{eqnarray}
and inflation occurs for $\epsilon < 1$. Self--consistency 
of the slow--roll approximation requires that $ {\rm max} 
\{ \epsilon , | \eta | \} \ll 1$.
The number of e--foldings of inflationary expansion 
that occur when the scalar field rolls from some 
value, $\phi$, to the value, $\phi_e$, corresponding to the end of inflation
is given by 
\begin{equation}
\label{efold}
N \equiv \ln a = \int^{t_e}_t dt \, H 
\approx -\frac{8\pi}{m_4^2} \int^{\phi_e}_{\phi}
\frac{V}{V'} \left( 1+\frac{V}{2\lambda} \right) d\phi   .
\end{equation}

The effect of the brane corrections is to 
enhance the value of the Hubble parameter 
relative to what it would be for a given pure Einstein
gravity model of the same energy density \cite{maartens}. 
This introduces
additional friction on the scalar field and further resists its motion 
down the potential, thereby enabling a steeper class of potentials 
to support inflation.
This is the basis behind the steep inflationary 
scenario \cite{steep}. The quadratic correction
relaxes the condition for slow--roll inflation in the RSII scenario
relative to the corresponding condition for the standard
model. Generically, steep inflation proceeds in the region of 
parameter space where 
$\rho \gg \lambda$ and naturally comes to an end when 
$\rho \approx \lambda$, since the conventional cosmological 
dynamics is recovered in this regime.

\subsection{Density Perturbations}

We now consider the generation of scalar and tensor 
perturbations in RSII inflation. Since many of the issues 
of perturbation theory in conventional inflationary 
models have already been covered in lectures at this school, we 
omit detailed discussions here and focus instead on the differences 
that arise between the two scenarios. 
We employ the normalization conventions of 
Ref. \cite{lidsey2}.  

We begin by recalling that the scalar perturbations 
generated during inflation that is driven by 
a single, self--interacting scalar field are {\em adiabatic}. 
The curvature perturbation on uniform density 
hypersurfaces is then given by $\zeta = H\delta \phi /\dot{\phi}$ 
and is determined by the scalar field fluctuation, $\delta \phi$, 
on spatially flat hypersurfaces \cite{perturbations}. 
Conservation of energy--momentum implies that $\zeta$ is conserved 
on large scales, a result that is independent of the 
specific form of the gravitational physics \cite{wands}. 
This implies that the 
amplitude of a mode when it re--enters the Hubble radius after inflation 
is related to the curvature perturbation by 
$A^2_S = 4 \langle \zeta^2 \rangle /25$, 
where the right--hand side is evaluated 
when the mode with comoving wavenumber, $k$, goes beyond the Hubble radius 
during inflation, i.e., when 
\begin{equation}
\label{kdef}
k(\phi) = a_e H(\phi ) \exp [ -N(\phi ) ]   ,
\end{equation}
where a subscript `e' denotes values at the 
end of inflation and $N =\int dt H (t)$ corresponds to the 
number of e--foldings of inflationary expansion that elapse 
between the time when the scale crosses the Hubble radius and 
the end of inflation [cf. Eq. (\ref{efold})]. 
Finally, the Gibbons--Hawking temperature of 
de Sitter space determines the magnitude of the field 
fluctuation, 
$\langle \delta \phi^2 \rangle = H^2/(4\pi^2)$, and 
we therefore deduce that the scalar perturbation amplitude has the form 
\begin{equation}
\label{scalar}
A^2_S = \left. \frac{1}{25\pi^2} \frac{H^4}{\dot{\phi}^2}
\right|_{k=aH}
\end{equation}
This is given in terms of the potential by \cite{maartens}
\begin{equation}
\label{scalaramp}
A^2_S = \frac{512\pi}{75m^6_4} \frac{V^3}{V'^2}
\left( 1+\frac{V}{2\lambda} \right)^3 
\end{equation}
after substitution of the Friedmann equation (\ref{branefriedmann}). 
We see that the amplitude is enhanced over that of the standard 
scenario by the bracketed term. 

\subsection{Gravitational Waves}

Although to first--order the gravitational waves decouple from the 
matter, the calculation of the tensor perturbation spectrum is more 
involved in braneworld cosmology because the perturbations extend 
into the bulk. In this subsection we review 
the method of Langlois, Maartens and Wands \cite{lmw}. 
To proceed analytically, it is necessary 
to assume pure de Sitter expansion on the brane and 
this is a good approximation 
if the inflation field is slowly rolling. 
It proves convenient to express the 
perturbed, five--dimensional metric in the form
\begin{equation}
ds^2_5 = {\cal{A}}^2 [-dt^2 + a^2 (\delta_{ij} +E_{ij})dx^i
dx^j] + dy^2 ,
\end{equation}
where $E_{ij}$ represents the perturbations. The warp factor is 
given by 
\begin{equation}
{\cal{A}} = (H/\alpha )
\sinh [\alpha (y_h -|y| )]   ,
\end{equation}
where the Cauchy horizons,  
$g_{00} (\pm y_h ) =0$, 
are located at $y=\pm y_h$, and the constant 
$\alpha =\kappa_4/\kappa_5 = (-\Lambda /6 )^{1/2}$ is determined by the 
bulk cosmological constant, $\Lambda$. Here and throughout, $\kappa_4^2 
\equiv 8\pi m^{-2}_4$ and $\kappa_5^2 \equiv 8\pi M^{-3}$.

The standard approach is to expand the metric perturbations as a 
Fourier series. In this case, and assuming that any anisotropic 
stresses are negligible, the linearly perturbed 
junction conditions (\ref{israeltwo}) reduce to
\begin{equation}
\label{pjunction}
\left. \frac{dE}{dy} \right|_{y=0} =0  ,
\end{equation}
where $E(t,y; \vec{k} )$ denotes the amplitude of the 
modes. Assuming a pure de Sitter expansion of the brane world--volume allows 
us to separate the corresponding gravitational wave equation 
of motion and then expand the amplitude into 
eigenmodes such that $E(t, y; \vec{k}) =
\int dm \varphi_m (t; \vec{k}) {\cal{E}}_m (y)$,
where $\varphi_m (t; \vec{k})$ and ${\cal{E}}_m (y)$
depend on the world--volume and bulk coordinates, respectively, and 
$m$ represents the separation constant. It can then be shown that 
the solution for the zero mode $(m=0)$ is determined 
in full generality up to a quadrature \cite{lmw}: 
\begin{equation}
\label{zerogeneral}
{\cal{E}}_0 =C_1 + C_2 \int^y dy' \frac{1}{{\cal{A}}^4(y')}  ,
\end{equation}
where $C_{1,2}$ are constants. In general, if a given mode 
diverges at the Cauchy horizon, it can not form part of the 
spectrum of orthonormal modes that constitute the basis 
of the Hilbert space for the quantum field. (Heuristically, this is because 
such a mode would produce an infinite contribution to 
the action and so it would cost too much energy to excite it). 
However, we must specify  $C_2=0$ to satisfy the boundary condition 
(\ref{pjunction}) and this removes the divergent part of the 
zero--mode. Thus, the physically relevant 
solution for the zero--mode is ${\cal{E}}_0 =C_1$. 
The non--zero modes are not excited -- modes 
where $m<3H/2$ remain divergent at the Cauchy horizon
even when Eq. (\ref{pjunction}) is satisfied and modes satisfying 
$m>3H/2$ remain in the vacuum state during inflation \cite{lmw}.

The zero--mode, $\varphi_0$, remains constant on super--Hubble radius 
scales, as in the four--dimensional scenario. The amplitude of the 
quantum fluctuation in this mode is then calculated by deriving an 
effective, five--dimensional 
action for the tensor perturbations and integrating over the 
fifth dimension. This results in a four--dimensional action 
that corresponds formally to a massless scalar field propagating 
in a FRW universe. The standard four--dimensional 
analysis may then be employed to determine the amplitude if
the action is normalized appropriately
when integrating over the fifth dimension. This 
requires that 
\begin{equation}
\label{Znormalization}
2\int_0^{y_h} dy C_1^2 {\cal{A}}^2 
=1
\end{equation}
and  implies that $C_1=\sqrt{\alpha} F(x )$, where 
\begin{equation}
\label{Fdefine}
\frac{1}{F^2} = \sqrt{1+x^2} -x^2 {\rm sinh}^{-1} 
\left( \frac{1}{x} \right)
\end{equation}
and $x \equiv H/\alpha$. Finally, the tensorial amplitude follows 
once each polarization state is interpreted as a quantum field 
propagating in a time--dependent potential \cite{lmw}: 
\begin{equation}
\label{symmetrictensor}
A_T^2   
= \left. \frac{\kappa_4^2}{50\pi^2}H^2F^2  \right|_{k=aH} .
\end{equation} 
The effects of the brane modifications are 
parametrized in terms of the `correction' 
function $F$. In the low--energy limit ($\rho \ll \lambda, 
x \ll 1)$, $F \approx 1$, whereas 
$F^2 \approx [27H^2m_4^2/(16\pi \lambda )]^{1/2}$ 
in the high--energy limit. 

\subsection{The Consistency Equation}

Since the scalar field slowly rolls down its potential, 
the amplitudes of the perturbations are not precisely 
scale--invariant. These variations are parametrized 
in terms of the spectral indices, or {\em tilts}, of the spectra and 
are defined by 
\begin{equation}
n_S\equiv 1 + d \ln A^2_S/d \ln k , \qquad 
n_T \equiv d\ln A^2_T /d \ln k
\end{equation}
for the scalar and tensor perturbations, respectively. 
The scalar spectral index may be expressed 
in terms of the slow--roll parameters: 
\begin{equation}
n_S - 1 = -6\epsilon +2\eta  .
\end{equation}
In the high energy limit $(\rho \gg \lambda , x \gg 1)$, the tilts
are given in terms of the potential and its first two derivatives by 
\begin{eqnarray}
\label{scalarindex}
n_S-1 \approx -\frac{m^2_4\lambda}{2\pi V}\left[ 3 
\frac{V'^2}{V^2} -\frac{V''}{V} \right] \\
\label{tensorindex}
n_T \approx -\frac{3m^2_4}{4\pi} \frac{\lambda V'^2}{V^3}  .
\end{eqnarray}

It is well known that since 
the scalar and tensor perturbations share a common origin
through the inflaton potential, $V(\phi)$, it is possible to 
relate them in a way that is independent 
of the functional form of the potential. (For a review, see, e.g., 
Ref. \cite{lidsey2}). This relationship 
is known as the 
{\em consistency equation} and, to lowest--order in the 
slow--roll approximation, determines the relative amplitudes of the tensor 
and scalar perturbations directly in terms of the tilt of the 
gravitational wave spectrum:  
\begin{equation}
\label{consistency}
\frac{A^2_T}{A^2_S} =  -\frac{1}{2} n_T  .
\end{equation}
 
Since it is independent of the potential, Eq. (\ref{consistency}) 
represents  a powerful 
test of single--field inflationary models and, in principle, 
failure to satisfy such a constraint could be employed to rule out 
such a class of models. At present,  the contribution of tensor 
perturbations to the large--angle CMB power spectrum is 
constrained to be no more than 30 $\%$ and, in practice, 
it will be very difficult, if not impossible, to 
measure the tilt of the tensor spectrum to a sufficient 
level of accuracy. Nevertheless, a 
cosmological background of gravitational waves could be detected 
through their contribution to the B--mode (curl) of the CMB 
polarization \cite{Bmode}
and interest is growing in this possibility 
in light of the recent detections of polarization 
in the CMB \cite{dasi,kogut}. 
In any case, even in the event that such a detection is not 
made, the consistency relation remains important
because it removes a free parameter 
(usually chosen to be $n_T$) 
when determining the best--fit models to the data. 

Given the importance of the consistency equation,  
it is clearly of interest to determine the form 
of the corresponding relations in braneworld cosmologies \cite{hl1,hl2}. 
We have seen that in the case of the RSII scenario,
the amplitudes of the perturbations are modified by the brane effects  
and these modifications 
become progressively more important at higher energy scales. 
Consequently, it 
is to be anticipated that the form of the consistency equation should reflect 
these differences. If so, this would provide a potentially observable 
test of RSII inflation. 

In the standard scenario, the consistency equation (\ref{consistency})
is derived by first differentiating the tensorial spectrum with 
respect to comoving wavenumber, $k$, and then 
relating a given scale to the corresponding value of the 
inflaton field through Eq. (\ref{kdef}). Any dependence 
on the first derivative of the inflaton potential may then be eliminated 
by substituting for the scalar perturbation amplitude and any remaining 
dependence on the inflaton potential itself may  
be removed by substituting for the tensor perturbation amplitude. 

In principle, an identical approach could be followed to derive the 
consistency equation in RSII inflation. However, given the complicated 
form of the amplitudes, this is algebraically very difficult (but not 
impossible) to accomplish. In view of this, we 
adopt a more elegant approach and proceed by 
defining a pair of new variables \cite{hl2}
\begin{eqnarray}
\label{defineb}
b \equiv \frac{1}{2} \sinh^{-1} x \\
\label{definebeta}
\beta \equiv \kappa_4 \frac{d \phi}{dN}  ,
\end{eqnarray}
where $x$ is defined after Eq. (\ref{Fdefine}). 
This implies that the Friedmann equation (\ref{branefriedmann}) 
and scalar field equation (\ref{branescalar}) 
reduce to a first--order, non--linear system 
of differential equations:
\begin{eqnarray}
\label{dotb}
\dot{b} =- \left( \frac{3\kappa^2_4}{8 \lambda} \right)^{1/2} \dot{\phi}^2
\\
\label{bprime}
\beta =-\left( \frac{8 \lambda}{3} \right)^{1/2} 
\frac{b'}{H}  .
\end{eqnarray}
Moreover, the correction function (\ref{Fdefine}) arising 
in the gravitational wave amplitude (\ref{symmetrictensor}) 
depends only on the single variable, $b$, and may be expressed in
terms of a single quadrature: 
\begin{equation}
\label{Fquad}
\frac{1}{F^2} = -4 \sinh^2 2b\int \frac{db}{\sinh^3 2b}   ,
\end{equation}
whereas the scalar perturbations (\ref{scalar}) 
depend on $\beta$:
\begin{equation}
\label{betascalar}
A_S^2 =\frac{\kappa^2_4}{25\pi^2} \frac{H^2}{\beta^2}   .
\end{equation}

We are now in a position to derive the form of the consistency 
equation in this scenario.
By employing the definitions (\ref{kdef}), 
(\ref{defineb}) and (\ref{definebeta}) 
and substituting Eqs.  
(\ref{symmetrictensor}), (\ref{bprime}) and (\ref{betascalar}) into 
Eq. (\ref{Fquad}), we find that 
\begin{equation}
\frac{1}{A^2_T} = 2 \int \frac{d \ln k}{A^2_S}   .
\end{equation}
Thus, differentiation with respect to comoving 
wavenumber recovers the consistency equation \cite{hl1,hl2}
\begin{equation}
\label{braneconsistency}
\frac{A^2_T}{A^2_S} =  -\frac{1}{2} n_T   .
\end{equation}

Remarkably, the form of the consistency equation  is {\em identical} 
to that of standard, single--field inflation. This is particularly 
surprising given that the gravitational physics 
is manifestly different in the two scenarios. 
Formally, this degeneracy between the consistency equations arises 
because the combination of observable parameters 
in Eq. (\ref{braneconsistency}) is independent of the 
brane tension, but it is not immediately transparent 
from Eqs. (\ref{scalaramp}), (\ref{Fdefine}) and (\ref{symmetrictensor}) 
why this should be so.  

\section{Asymmetric Braneworld Inflation}

\begin{figure}
\centering
\includegraphics[height=4cm]{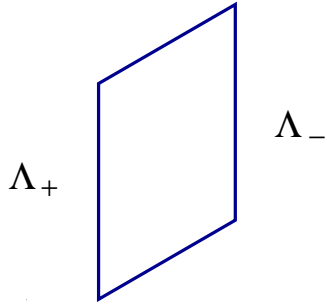}
\caption{In the asymmetric RSII braneworld scenario, there is 
no reflection symmetry imposed on the bulk dimension and the 
cosmological constant may take different values on either side of the 
brane.}
\label{mexico2}      
\end{figure}

We now proceed in this Section to
consider an extension of the RSII scenario
where the $Z_2$ reflection symmetry in the bulk dimension 
is no longer imposed. This implies that the brane may be embedded in 
five--dimensional AdS space where the value of the 
cosmological constant differs on either side of the brane (Fig. 
\ref{mexico2}). Similar analyses to those summarized in Sections 
3 and 4 may be followed to derive the Friedmann 
equation and the inflationary perturbation spectra. 
Here, we omit many of the details and simply highlight the main results. 

The spatial components of the junction conditions reduce to 
\cite{othermoves,stw,ch}
\begin{equation}
\label{junction+-}
\left( \alpha_+ +H^2 \right)^{1/2} +\left( 
\alpha_- +H^2 \right)^{1/2} =
\frac{\kappa^2_5 \rho}{3}      ,
\end{equation}
where 
\begin{equation}
\alpha_{\pm} \equiv -\kappa^2_5 \Lambda_{\pm}/6
\end{equation}
and $\Lambda_{\pm}$ are the bulk cosmological constants 
either side of the brane. 
Dimensional reduction relates the four-- and five--dimensional 
Newton constants: 
\begin{equation}
\frac{\kappa^2_5}{\kappa^2_4} = \frac{1}{2} 
\left( \frac{1}{\sqrt{\alpha}_+} +\frac{1}{\sqrt{\alpha_-}} 
\right)   .  
\end{equation}

Solving Eq. (\ref{junction+-}) yields the Friedmann equation
\cite{othermoves,stw,ch}: 
\begin{equation}
\label{friedmann+-}
H^2 =\frac{\kappa_5^4 \rho^2}{36} - \frac{1}{2} \left( \alpha_- +\alpha_+ 
\right) +\frac{9}{4 \kappa^4_5 \rho^2} \left( \alpha_- -\alpha_+ 
\right)^2  
\end{equation}
and, as in the symmetric scenario, a brane tension may be introduced in order 
to recover the required linear dependence on the energy density 
at low energy scales. Relaxing the $Z_2$ 
symmetry results in the appearance of the third term on the right--hand 
side of Eq. (\ref{friedmann+-}). Note how this term is proportional 
to $\rho^{-2}$ and becomes negligible at high 
energies. (However, this does not imply that 
such a term diverges at very low energies, 
since the density as shown 
here represents the total energy density on the brane. 
This consists of both the matter contributions as well as 
the brane tension, $\lambda$). 
It is worth remarking that the Friedmann 
equation (\ref{friedmann+-}) exhibits an infra--red/ultra--violet duality, 
in the sense that it is invariant under the 
transformation of the energy density, $\kappa^2_5 \rho \leftrightarrow 
9| \alpha_+ -\alpha_- |/(\kappa_5^2\rho)$.

Since the bulk space on either side of the brane 
is AdS, the Gauss--Codazzi equation (\ref{codazzi}) once more implies that 
energy--momentum is conserved on the brane and Eq. (\ref{fluid}) 
therefore remains valid. This is important because it implies that 
the argument of Section 4.2 may be employed once more 
to determine the amplitude of the scalar perturbations. 
Consequently, the amplitude is given by 
Eq. (\ref{scalar}), although its specific dependence on the inflaton potential 
is altered from that of the symmetric scenario
due to the additional term arising 
in the Friedmann equation (\ref{friedmann+-}). 
Indeed, substituting Eq. 
(\ref{friedmann+-}) and the scalar field equation 
(\ref{branescalar}) into Eq. (\ref{scalar}) implies that \cite{hl2}
\begin{eqnarray}
\label{scalar1}
A_S^2 =  \frac{9}{25\pi^2} \frac{1}{V'^2} 
\left[  
\frac{\kappa_5^4 (V + \lambda )^2}{36} - \frac{1}{2} 
\left( \alpha_- +\alpha_+ 
\right)  \right. \nonumber \\ \left. 
\left. +\frac{9}{4 \kappa^4_5 (V + \lambda )^2} \left( \alpha_- -\alpha_+ 
\right)^2
\right]^3 \right|_{k=aH}
\end{eqnarray}

The tensor spectrum for this model has been calculated in Ref. \cite{hl2}
by extending the method of Ref. \cite{lmw}. The result is 
\begin{equation}
\label{tensor+-}
\left. A^2_T =\frac{\kappa^2_5}{50\pi^2} H^2 J^2
\right|_{k=aH}
\end{equation}
where 
\begin{equation}
\label{Jdefine2}
\frac{1}{J^2} =  \frac{1}{2\sqrt{\alpha_-} F^2(x_-)} 
+ \frac{1}{2\sqrt{\alpha_+}F^2(x_+)}   ,
\end{equation}
the functional form of $F=F(x_{\pm})$ is given by Eq. 
(\ref{Fdefine}) and 
$x_{\pm} \equiv H/\sqrt{\alpha_{\pm}}$.
At low energy scales ($x_{\pm} \ll 1$), we find that 
$J \rightarrow \kappa_4/\kappa_5$, implying that the standard expression 
is recovered in this limit, as expected. 

The consistency equation can be derived in this model \cite{hl2}.  
Given the results of Section 4, the simplest approach is 
to assume {\em a priori} that the consistency equation 
has the same form as that of Eq. (\ref{consistency}) and to then verify 
that this is indeed the case. Let us therefore substitute the scalar 
and tensor perturbations, 
Eqs. (\ref{scalar}) and (\ref{tensor+-}), directly into 
Eq. (\ref{consistency}). We find that 
\begin{equation}
\label{require1}
\frac{d \ln (HJ)}{d H} \frac{dH}{dN} = - \frac{\kappa^2_5}{18}
\frac{J^2V'^2}{H^4}   ,
\end{equation}
where we have employed the slow--roll approximation in the form 
$d\ln k \approx  d \ln a = dN$. Noting that the scalar field equation, 
$3H\dot{\phi} = -V'$, may be expressed in the somewhat unconventional form
\begin{equation}
\label{require2}
\frac{dH}{dN} = - \frac{d H}{dV} \frac{V'^2}{3H^2}  
\end{equation}
then allows us to simplify Eq. (\ref{require1}) 
through substitution of Eq. (\ref{require2}) \cite{hl2}: 
\begin{equation}
\label{require3}
H^4 \frac{dH}{dV} \frac{d (HJ)^{-2}}{dH}  = - \frac{\kappa_5^2}{3} .
\end{equation}
Eq. (\ref{require3}) represents a {\em necessary} 
condition for the consistency equation (\ref{consistency}) to hold. 
It should be emphasized that this condition
applies to any (single field) braneworld 
scenario, where energy--momentum is conserved on the 
brane. In general, a
given  braneworld model may be characterized by the functional 
form of its Friedmann equation, i.e., by the 
dependence of the Hubble parameter on the inflaton potential, 
$H=H(V)$. Once this relation has been established, 
Eq. (\ref{require3}) may then be interpreted as 
a constraint that must be satisfied by the correction function, $J=J(H)$,
if the consistency equation 
is to remain degenerate. To illustrate this, consider the standard 
scenario, where $H \propto V^{1/2}$. We see immediately 
that the left--hand side of Eq. (\ref{require3}) 
is constant when $J=1$.

It is possible, after some lengthy algebra, to confirm that 
Eq. (\ref{require3}) is indeed satisfied for the asymmetric RSII scenario 
when the correction 
to the tensor spectrum takes the form given by Eq. (\ref{Jdefine2})
and we therefore conclude that even in this generalized, 
asymmetric RSII model, the consistency equation remains degenerate \cite{hl2}: 
\begin{equation}
\frac{A^2_T}{A^2_S} =  -\frac{1}{2} n_T  .
\end{equation}

Before concluding this Section, we briefly 
consider the braneworld model of Ref. \cite{gklr}, 
where the brane is embedded in a five--dimensional 
bulk space with a stabilized radius. This model differs from the RSII scenario 
in that the Friedmann equation, and therefore the scalar perturbation 
spectrum, remain unmodified. 
However, the presence of the fifth dimension becomes apparent 
through a correction to the tensor spectrum. Thus, one would certainly 
expect the consistency equation to be modified in this 
model. However, surprisingly, this is not the case -- the correction 
to the gravitational wave amplitude 
is given by $J^2=(1-\alpha H^2)^{-2}$, where $\alpha$ is a constant 
with a numerical value determined by the radius of the extra dimension
\cite{gklr}. 
One may readily deduce that a correction of this form satisfies 
Eq. (\ref{require3}) and, moreover, represents the general 
solution to Eq. (\ref{require3}) when the Friedmann equation has the 
standard form. Thus, we now know of four classes of 
inflationary cosmologies, modelled on different gravitational physics, 
where the consistency equation remains robust. 

Such a degeneracy implies that the task of 
identifying the correct inflationary model through 
observations will be more difficult. 
On the other hand, although the consistency equation may be interpreted 
as a prediction of single field inflation, it should be emphasized that it is 
a prediction relating the {\em primordial} perturbations. 
In particular, in the above analyses we have neglected the influence of the 
bulk space on the subsequent evolution of the perturbations. 
This is equivalent to assuming that the projection of the 
five--dimensional Weyl tensor vanishes to linear order. More generally, 
however, the backreaction of the bulk will perturb the 
bulk space away from conformal invariance and generate a non--trivial Weyl 
tensor in five dimensions. 
This results in a non--local energy--momentum 
source in the gravitational field equations
when projected down to four dimensions \cite{gordonmaartens}. 
As a result, the background dynamics is altered. The subsequent 
evolution of the perturbations is difficult to determine 
in general, because the system of equations is not closed, 
although it is expected that it will be model--dependent to some extent. 

A further assumption that we have made is that the field 
is rolling sufficiently slowly down its potential. This assumption 
can be relaxed in the standard scenario by 
working to the `next--to--leading' order in the slow--roll
approximation. In this regime, it has been shown that the consistency 
equation (\ref{consistency}) receives modifications \cite{ckll}: 
\begin{equation}
\label{hc}
n_T = -2 \frac{A^2_T}{A^2_S} \left[ 1- \frac{A^2_T}{A^2_S} 
+ (1-n_S) \right]   .
\end{equation}
The question that naturally arises, therefore, is 
whether the corresponding consistency equations in the RSII 
scenarios receive similar corrections or whether the 
degeneracy can be lifted by moving away from the 
slow--roll approximation. The answer to this and related questions 
must be left for future work.

\section{Gauss--Bonnet Braneworld Cosmology}

As well as developing the framework for testing braneworld 
inflation through a confrontation with observations, another 
important task is to enhance the connection of the 
scenario with string/M--theory. One approach towards this goal 
is to include combinations of higher--order curvature invariants 
in the bulk action \cite{HD,GBrefs,onlylocal,gbf,davis,gw,bcdd,ln}.  
Within the context of the 
AdS/CFT correspondence, such terms arise as next--to--leading 
order corrections in the $1/N$ expansion 
of the CFT \cite{largeN}. The Gauss--Bonnet combination, 
$\hat{R}^2 -4\hat{R}_{ab}\hat{R}^{ab} 
+\hat{R}_{abcd}\hat{R}^{abcd}$, is of 
particular relevance, given that it is 
the unique combination in five dimensions 
that results in second--order field 
equations in the metric and it also appears as a 
leading--order quantum correction in the heterotic string theory action
\cite{stringGB,bd,d86}. 

The extension of the RSII scenario to include this term is presently 
attracting attention. The five--dimensional field equations admit 
Schwarzschild--AdS 
space as a solution \cite{bd,cai}
and the Friedmann equation may be derived 
through a variety of methods. These include generalizing Birkhoff's theorem 
\cite{gbf}, varying the boundary terms in the action \cite{davis}, 
or by employing the 
formalism of differential forms \cite{gw}. 
When the bulk space is $Z_2$ symmetric, 
the Friedmann equation takes the form \cite{gbf,davis,gw}
\begin{equation}
\label{friedmannGB}
H^2= \frac{c_+ +c_- -2}{8\alpha} \,,
\end{equation}
where
\begin{equation}
\label{defc}
c_{\pm} = \left\{ \left[ \left( 1+\frac{4}{3}\alpha \Lambda \right)^{3/2} 
+ \frac{\alpha}{2} \kappa_5^4 \rho^2 \right]^{1/2} 
\pm \sqrt{\frac{\alpha}{2}} \kappa_5^2 \rho \right\}^{2/3}   ,
\end{equation}
the bulk cosmological constant is $\Lambda$ 
and $\alpha> 0 $ represents the Gauss--Bonnet coupling constant. 
As in the models discussed 
above, conservation of energy--momentum on the brane follows 
directly from the Gauss--Codazzi equations. 

Despite the rather complicated form of Eq. (\ref{friedmannGB}), it is 
possible to make progress analytically by introducing 
a new variable, $r$ \cite{ln}: 
\begin{equation}
\label{defr}
\rho \equiv \left( \frac{2b}{\alpha \kappa^4_5} \right)^{1/2}
\sinh r 
\end{equation}
and defining the constant 
\begin{equation}
\label{defb}
b \equiv \left( 1+\frac{4}{3} \alpha \Lambda \right)^{3/2} \,.
\end{equation}
Substituting Eqs. (\ref{defr}) and (\ref{defb}) 
into Eq.~(\ref{defc}) then implies that
\begin{equation}
c_{\pm} = b^{1/3} \exp(\pm 2r/3) 
\end{equation}
and, as a result, 
the Friedmann equation (\ref{friedmannGB}) simplifies considerably \cite{ln}: 
\begin{equation}
\label{friedmannrGB}
H^2 = \frac{1}{4\alpha} \left[ b^{1/3} \cosh \left( \frac{2r}{3} \right) 
-1 \right] \,.
\end{equation}

It may be verified that the Friedmann equation (\ref{friedmannrGB})
exhibits a quadratic dependence on the total energy density 
in the low energy limit corresponding to the RSII model. 
At sufficiently high energies, however, 
the dependence scales as $H^2 \propto \rho^{2/3}$. 
The condition for inflation to proceed in this regime 
is simply that the pressure of the matter be negative, $p<0$. 
More precisely, slow--roll parameters may be introduced
and, in the slow--roll 
limit where the inflaton potential dominates the brane tension, $\lambda$,
they are given by \cite{ln} 
\begin{eqnarray}
\label{epsilonGB}
\epsilon &=& \left( \frac{2\lambda}{\kappa_4^2}~\frac{V'^2}{V^3} \right) 
\left[ \frac{2b^{2/3}}{27}~ 
\frac{\sinh (2r/3) \, \tanh r \, \sinh^2 r}{\left[ 
b^{1/3} \cosh (2r/3) -1 \right]^2} \right] \,, \\
\label{etaGB}
\eta &=& \left( \frac{2\lambda}{\kappa_4^2}~ \frac{V''}{V^2} \right) 
\left[ \frac{2b^{1/3}}{9}~ 
\frac{\sinh^2 r}{b^{1/3} \cosh (2r/3) -1} \right] .
\end{eqnarray}
The terms in the square brackets parametrize 
the effects of the 
Gauss--Bonnet contribution. These are monotonically decreasing functions 
of $r$ and tend to unity from above as $\{ r, \alpha \}
\rightarrow 0$. In this limit, the slow--roll parameters reduce to those 
of the RSII model. It follows, therefore, that the Gauss--Bonnet  
contribution tightens the constraints for inflation to proceed 
relative to the RSII scenario. 

A further consequence of introducing a Gauss--Bonnet term is that 
the spectrum of perturbations is altered. For example, 
in the high--energy limit where $H^2 \propto \rho^{2/3}$, 
we find that $A_S^2 \propto H^4/\dot{\phi}^2 \propto 
H^6/V'^2 \propto (V/V')^2$. Thus, for the case of an exponential
potential, the spectrum is pushed very close 
to a scale--invariant form \cite{ln}. 
This is interesting given that potentials 
of this nature generically arises in a number of particle physics 
inspired settings.

Finally, it would be of interest to determine whether 
the degeneracy of the inflationary consistency equation 
is lifted by introducing a Gauss--Bonnet term into the bulk action. 
To date, the gravitational wave spectrum in this model 
has yet to be determined. The calculation of the spectrum is more 
involved than that of the RSII scenario, because the 
linearly perturbed junction conditions must be employed to impose the
necessary boundary conditions on the perturbations.  

\section{Concluding Remark}

To summarize, inflationary cosmology based on Randall--Sundrum 
braneworlds remains a rich environment for 
future work. The scenario has already revealed unexpected surprises 
and surely has more to offer. 
It is well motivated from a string theoretic 
perspective in view of its 
close relationship with the AdS/CFT correspondence.
It is sufficiently simple to provide a framework for 
performing analytical calculations
and thereby making observational predictions. 
Thus, it provides a unique window into higher--dimensional physics.

\vspace{.3in}
JEL is supported by the Royal Society. It is a pleasure
to thank G. Huey 
and N. Nunes with whom various results presented in Sections 4 \& 5 and 
Section 6 were derived.


\begin{thebibliography}{99.}

\bibitem{temp}
C. L. Bennett {\em  et al.}, astro-ph/0302207; 
G. Hinshaw {\em et al.}, astro-ph/0302217.

\bibitem{wmap} 
D. N. Spergel {\em et al.}, astro-ph/0302209.

\bibitem{Perlmutter}
B. P. Schmidt {\it et al.},
Astrophys. J.  {\bf 507}, 46 (1998); 
A. G. Riess {\it et al.}  [Supernova Search Team Collaboration],
Astron. J.  {\bf 116}, 1009 (1998); 
S. Perlmutter {\it et al.}  [Supernova Cosmology Project Collaboration],
Astrophys. J.  {\bf 517}, 565 (1999). 

\bibitem{simplest}
A. A. Starobinsky, Phys. Lett. {\bf 91B}, 99 (1980); 
A. H. Guth, Phys. Rev. {\bf D23}, 347 (1981); A. Albrecht 
and P. J. Steinhardt, Phys. Rev. Lett. {\bf 48}, 
1220 (1982); S. W. Hawking and I. G. Moss, Phys. Lett. {\bf 110B}, 
35 (1982); A. D. Linde, Phys. Lett. {\bf 108B}, 389 (1982); 
A. D. Linde, Phys. Lett. {\bf 129B}, 177 (1983). 

\bibitem{lidlyth} 
A. R. Liddle and D. H. Lyth, 
{\em Cosmological Inflation and Large-Scale Structure}
(Cambridge University Press, Cambridge, 2000).

\bibitem{perturbations}
V. Mukhanov and G. Chibisov, Pis'ma Zh. Eksp. Teor. Fiz. {\bf 33}, 
549 (1981) [JETP Lett. {\bf 33}, 532 (1981), astro-ph/0303077]; 
A. H. Guth and S. Y. Pi, Phys. Rev. Lett. {\bf 49}, 1110 (1982);
S. W. Hawking, Phys. Lett.  {\bf 115B}, 295 (1982);
A. D. Linde, Phys. Lett. {\bf 116B}, 335 (1982);
A. A. Starobinsky, Phys. Lett. {\bf 117B}, 175 (1982); 
A. A. Starobinsky, Sov. Astron. Lett. {\bf 9}, 302 (1983); 
J. M. Bardeen, P. J. Steinhardt, and M. S. Turner,
Phys. Rev. {\bf D28}, 679 (1983); D. H. Lyth, Phys. Rev. 
{\bf D31}, 1792 (1985). 

\bibitem{wmap1}
H. V. Peiris {\em et al.}, astro-ph/0302225.

\bibitem{kogut} A. Kogut {\em et al.}, astro-ph/0302213.

\bibitem{komatsu}
E. Komatsu {\em et al.}, astro-ph/0302223.

\bibitem{bridle}
S. L. Bridle, A. M. Lewis, J. Weller, and G. Efstathiou, 
astro-ph/0302306. 


\bibitem{kkmr} S. Dodelson and L. Hui, 
astro-ph/0305113; W. H. Kinney, E. W. Kolb, 
A. Melchiorri, and A. Riotto, astro-ph/0305130; A. R. Liddle and 
S. M. Leach, astro-ph/0305263. 

\bibitem{white} 
W. Hu and M. White, Astrophys. J. {\bf 479}, 568 (1997); 
D. N. Spergel and M. Zaldarriaga, Phys. Rev. Lett. {\bf 79}, 2180 (1997). 

\bibitem{string}  J. Polchinski, {\em String Theory} 
(Cambridge University Press, Cambridge, 1998). 

\bibitem{Dbrane} C. V. Johnson, {\em D--Branes} 
(Cambridge University Press, Cambridge, 2003). 

\bibitem{lwc}
J. E. Lidsey, D. Wands, and E. J. Copeland, Phys. Rep. 
{\bf 337}, 343 (2000); M. Quevedo, hep-th/0210292; 
M. Gasperini and G. 
Veneziano, Phys. Rep. {\bf 373}, 1 (2003).

\bibitem{witten} E. Witten, Nucl. Phys. {\bf B443}, 85 (1995). 

\bibitem{townsend} P. Townsend, Phys. Lett. {\bf B350}, 184 (1995). 

\bibitem{polchinski} J. Polchinski, Phys. Rev. Lett. {\bf 75}, 4724 (1995). 

\bibitem{early} K. Akama, 
Lect. Notes Phys. {\bf 176}, 267 (1982), hep-th/0001113; 
V. A. Rubakov and M. E. Shaposhnikov, 
Phys. Lett. {\bf 125B}, 136 (1983); 
M. Visser, Phys. Lett. {\bf 159B}, 22 (1985).

\bibitem{hw}
P. Ho\v{r}ava and E. Witten, Nucl. Phys. {\bf B460}, 506 (1996); 
P. Ho\v{r}ava and E. Witten, Nucl. Phys. {\bf B475}, 94 (1996).

\bibitem{lukas}
A. Lukas, B. A. Ovrut, K. S. Stelle, and D. Waldram, 
Phys. Rev. {\bf D59}, 086001 (1999).

\bibitem{hwcos} 
K. Benakli, Int. J. Mod. Phys. {\bf D8}, 153 (1999); 
K. Benakli, Phys. Lett. {\bf B447},  51 (1999); 
A Lukas, B. A. Ovrut, and D. Waldram, Phys. Rev. {\bf D60},  086001 (1999); 
H. S. Reall, Phys. Rev. {\bf D59}, 103506 (1999); 
J. E. Lidsey, Class. Quantum Grav. {\bf 17}, L39 (2000). 

\bibitem{ven} M. Gasperini and G. Veneziano, Astropart. Phys. 
{\bf 1}, 317 (1991). 

\bibitem{turok} 
J. Khoury, B. A. Ovrut, P. J. Steinhardt, and N. Turok, 
Phys. Rev. {\bf D64}, 123522 (2001). 

\bibitem{hier}
N. Arkani--Hamed, S. Dimopoulos, and G. Dvali, 
Phys. Lett. {\bf B429}, 263 (1998); 
I. Antoniadis, N. Arkani--Hamed, S. Dimopoulos, and G. Dvali, 
Phys. Lett. {\bf B436}, 257 (1998).

\bibitem{RSI}
L. Randall and R. Sundrum, Phys. Rev. Lett. {\bf 83}, 3370 (1999).

\bibitem{RSII} L. Randall and R. Sundrum, Phys. Rev. Lett. {\bf 83}, 
4690 (1999). 

\bibitem{holographic} G. `t Hooft, gr-qc/9310026; 
L. Susskind, {\sl J. Math. Phys.} {\bf 36}, 6337 (1995). 

\bibitem{adscft}
J. M. Maldacena,
Adv. Theor. Math. Phys. {\bf 2}, 231 (1998); 
E. Witten, Adv. Theor. Math. Phys. {\bf 2}, 505 (1998); 
S. Gubser, I. Klebanov, and A. Polyakov, Phys. Lett. {\bf B428}, 105 (1998);  
O. Aharony, S. Gubser, J. Maldacena, H. Ooguri, and Y. Oz,
Phys. Rep. {\bf 323}, 183 (2000).

\bibitem{hrr}
S. W. Hawking, T. Hertog, and H. S. Reall,
Phys. Rev. {\bf D62} 043501 (2000); 
S. Nojiri, S. D. Odintsov, and S. Zerbini,
Phys. Rev. {\bf D62} 064006 (2000); 
S. Nojiri and S. Odintsov,
Phys. Lett. {\bf B484}, 119 (2000); 
L. Anchordoqui, C. Nunez, and K. Olsen,
J. High Energy Phys. {\bf 10}, 050 (2000); 
S. Nojiri and S. Odintsov,
Phys. Lett. {\bf B494}, 135 (2000); 
S. Gubser,
Phys. Rev. {\bf D63}, 084017 (2001); 
T. Shiromizu and D. Ida,
Phys. Rev. {\bf D64}, 044015 (2001). 

\bibitem{ver} E. Verlinde, hep-th/0008140; I. Savonije and 
E. Verlinde, Phys. Lett.  {\bf B507}, 305 (2001).

\bibitem{argurio} R. Argurio, hep-th/9807171; J. P. Gauntlett, hep-th/9705011.

\bibitem{lidsey1}
J. E. Lidsey, Phys. Rev. {\bf D64}, 063507 (2001). 

\bibitem{othermoves}
P. Kraus, J. High Energy Phys. {\bf 12}, 011 (1999); 
D. Ida,  J. High Energy Phys. {\bf 09}, 014 (2000). 

\bibitem{reall} 
H. A. Chamblin and H. S. Reall, Nucl. Phys. {\bf B562}, 133 (1999); 
H. A. Chamblin, M. J. Perry, and H. S. Reall, J. High Energy Phys. 
{\bf 09}, 014 (1999). 

\bibitem{bv}
C. Barcelo and M. Visser, Phys. Lett. {\bf B482}, 183 (2000).

\bibitem{israel} 
W. Israel, Nuovo Cim. {\bf 44B}, 1 (1966). 

\bibitem{bine} P. Bin\'etruy, 
C. Deffayet, and D. Langlois, Nucl. Phys. 
{\bf B565}, 269 (2000); 
P. Bin\'etruy, C. Deffayet, U. Ellwanger, and 
D. Langlois, Phys. Lett. {\bf B477}, 285 (2000). 

\bibitem{flanagan} E. E. Flanagan, S. -H. Tye, and 
I. Wasserman, Phys. Rev. {\bf D62}, 044039 (2000). 

\bibitem{4a} J. M. Cline, C. Grojean, and G. Servant, 
Phys. Rev. Lett. {\bf 83}, 4245 (1999); C. C\'saki, M. Graesser, C. 
Kolda, and J. Terning, Phys. Lett. {\bf B462}, 34 (1999).

\bibitem{shiromizu} T. Shiromizu, K. Maeda, and M. Sasaki, 
Phys. Rev. {\bf D62}, 024012 (2000).  

\bibitem{maartens} R. Maartens, D. Wands, 
B. Bassett, and I. Heard, Phys. Rev. {\bf D62}, 041301 (2000).

\bibitem{steep}
E. J. Copeland, A. R. Liddle, and 
J. E. Lidsey, Phys. Rev. {\bf D64}, 023509 (2001).

\bibitem{lidsey2} J. E. Lidsey, A. R. Liddle, E. W. Kolb, E. J. Copeland, 
T. Barreiro, and M. Abney, Rev. Mod. Phys. {\bf 69}, 373 (1997).

\bibitem{wands}
D. Wands, K. A. Malik, D. H. Lyth, and A. R. Liddle, 
Phys. Rev. {\bf D62}, 043527 (2000).

\bibitem{lmw} D. Langlois, R. Maartens, and D. Wands, Phys. Lett. 
{\bf B489}, 259 (2000).

\bibitem{dasi} J. Kovac {\em et al.}, Nat. {\bf 420}, 772 (2002).

\bibitem{Bmode}
M. Kamionkowski, A. Kosowsky, and A. Stebbins, 
Phys. Rev. Lett. {\bf 78}, 2058 (1997); 
U. Seljak and M. Zaldarriaga, Phys. Rev. Lett. 
{\bf 78}, 2054 (1997). 

\bibitem{hl1} G. Huey and J. E. Lidsey, 
Phys. Lett. {\bf B514}, 217 (2001). 

\bibitem{hl2} G. Huey and J. E. Lidsey, 
Phys. Rev. {\bf D66}, 043514 (2002). 

\bibitem{stw}
H. Stoica, S. -H. Tye, and I. Wasserman, 
Phys. Lett. {\bf B482}, 205 (2000).

\bibitem{ch}
H. Collins and B. Holdom, Phys. Rev. {\bf D62}, 105009 (2000); 
N. Deruelle and T. Dolezel, 
Phys. Rev.  {\bf D62}, 103502 (2000); 
P. Bowcock, C. Charmousis, and R. Gregory, Class. 
Quantum Grav. {\bf 17}, 4745 (2000); 
W. B. Perkins, Phys. Lett. {\bf B504}, 28 (2001). 

\bibitem{gklr}
G. F. Giudice, E. W. Kolb, J. Lesgourgues, and A. Riotto, Phys. Rev. 
{\bf D66}, 083512 (2002).

\bibitem{gordonmaartens}
C. Gordon and R. Maartens, Phys. Rev. {\bf D63}, 044022 (2001).

\bibitem{ckll}
E. J. Copeland, E. W. Kolb, A. R. Liddle, and J. E. Lidsey, 
Phys. Rev. {\bf D49}, 1840 (1994). 

\bibitem{HD}
S. Nojiri and S. D. Odintsov, 
J. High Energy Phys. {\bf 07}, 049  (2000); 
M. Giovannini, Phys. Rev. {\bf D63}, 064011 (2001); 
S. Mukohyama, Phys. Rev.  {\bf D63}, 104025 (2001); 
G. Kofinas, J. High Energy Phys. {\bf 08}, 034 (2001); 
S. Nojiri, S. D. Odintsov, and S. Ogushi,
Int. J. Mod. Phys. {\bf A16}, 5085 (2001); 
S. Nojiri, S. D. Odintsov, and S. Ogushi,
Phys. Rev. {\bf D65}, 023521 (2002). 

\bibitem{GBrefs}
J. E. Kim, B. Kyae, and H. M. Lee, Phys. Rev. {\bf D62}, 045013 (2000); 
N. Deruelle and T. Dolezel, Phys. Rev. {\bf D62}, 103502  (2000); 
I. Low and A. Zee, Nucl. Phys. {\bf B585}, 395  (2000); 
O. Corradini and Z. Kakushadze, Phys. Lett. {\bf B494}, 302 (2000); 
J. E. Kim, B. Kyae, and H. M. Lee, 
Nucl. Phys. {\bf B582}, 296 (2000); 
Erratum--{\em ibid}  {\bf B591}, 587 (2000); 
J. E. Kim and H. M. Lee, Nucl. Phys. {\bf B602}, 346 (2001); 
B. Abdesselam and N. Mohammedi, 
Phys. Rev. {\bf D65}, 084018 (2002); C. Germani and C. F. Sopuerta, 
Phys. Rev. Lett. {\bf 88}, 231101 (2002); J. E. Lidsey, S. Nojiri, and
S. D. Odintsov, J. High Energy Phys. {\bf 06}, 026 (2002). 

\bibitem{onlylocal}
N. E. Mavromatos and J. Rizos, Phys. Rev. {\bf D62}, 124004 (2000); 
I. P. Neupane, J. High Energy Phys. {\bf 09}, 040 (2000); 
I. P. Neupane, Phys. Lett. {\bf B512}, 137 (2001); 
K. A. Meissner and M. Olechowski, Phys. Rev. Lett. {\bf 86}, 3708 
(2001); 
Y. M. Cho, I. Neupane, and P. S. Wesson, 
Nucl. Phys. {\bf B621}, 388 (2002).  
 
\bibitem{gbf}
C. Charmousis and J. Dufaux, Class. Quantum Grav. {\bf 19}, 4671 (2002). 

\bibitem{davis} 
S. C. Davis, Phys. Rev. {\bf D67}, 024030 (2003). 

\bibitem{gw}
E. Gravanis and S. Willison, 
hep-th/0209076. 

\bibitem{bcdd}
P. Binetruy, C. Charmousis, S. C. Davis, and J. Dufaux, 
Phys. Lett. {\bf B544}, 183 (2002).

\bibitem{ln} J. E. Lidsey and N. Nunes, astro-ph/0303168.

\bibitem{largeN}
A. Fayyazuddin and M. Spalinski, 
Nucl. Phys. {\bf B535}, 219 (1998); 
O. Aharony, A. Fayyazuddin, and J. Maldacena, 
J. High Energy Phys. {\bf 07}, 013 (1998).

\bibitem{stringGB}
B. Zwiebach, 
Phys. Lett. {\bf 156B}, 315 (1985); 
A. Sen, Phys. Rev. Lett. {\bf 55}, 1846 (1985);
R. R. Metsaev and A. A. Tseytlin, 
Nucl. Phys. {\bf B293}, 385 (1987).

\bibitem{bd}
D. G. Boulware and S. Deser, 
Phys. Rev. Lett. {\bf 55}, 2656 (1985).

\bibitem{d86}
N. Deruelle and J. Madore, 
Mod. Phys. Lett. {\bf A1}, 237 (1986); 
N. Deruelle and L. Farina--Busto, Phys. Rev.  {\bf D41}, 3696 (1990).

\bibitem{cai}
R. G. Cai, Phys. Rev. {\bf D65}, 084014 (2002). 

\end{thebibliography}
\end{document}